\documentclass[prb,superscriptaddress,showpacs]{revtex4-1}
\usepackage{graphicx}
\usepackage{dcolumn}
\usepackage{bm}
\usepackage{amsmath}
\usepackage{amssymb}
\usepackage{ulem}
\usepackage{color}
\usepackage{float}

\newcommand{\ket}[1]{\left|#1\right>}
\newcommand{\bra}[1]{\left<#1\right|}
\newcommand{\braket}[2]{\left<#1|#2\right>}

\def\upuparrows{\uparrow\!\uparrow}
\def\updownarrows{\uparrow\!\downarrow}
\def\downdownarrows{\downarrow\!\downarrow}

\def\B{{\mathcal B}}
\def\QM{{Q_{\rm min}}}
\def\NL{{M_\Lambda^{}}}

\def\Tr{{\bf Tr}}
\def\bk{{\bf k}}
\def\bq{{\bf q}}
\def\br{{\bf r}}
\def\bQ{{\bf Q}}
\def\bL{{\bf L}}
\def\HT{Hex$^{(2)}$}
\def\HH{Hex$^{(4)}$}

\def\uo{\underline{1}}
\def\ut{\underline{2}}

\begin{document}

\title{Hartree-Fock Ground State Phase Diagram of Jellium}
\author{L. Baguet}
\affiliation{LPTMC, UMR 7600 of CNRS, Universit\'e P. et M. Curie, Paris, France}
\author{F. Delyon}
\affiliation{CPHT, UMR 7644 of CNRS, \'Ecole Polytechnique, Palaiseau, France}
\author{B. Bernu}
\affiliation{LPTMC, UMR 7600 of CNRS, Universit\'e P. et M. Curie, Paris, France}
\author{M. Holzmann}
\affiliation{LPTMC, UMR 7600 of CNRS, Universit\'e P. et M. Curie, Paris, France}
\affiliation{Univ. Grenoble 1/CNRS, LPMMC UMR 5493, Maison des Magist\`{e}res, 38042 Grenoble, France}

\date{\today}
\begin{abstract}
We calculate the ground state phase diagram of the homogeneous electron gas in three dimensions within the Hartree-Fock approximation and show that broken symmetry states are energetically favored at any density against the homogeneous Fermi gas state with isotropic Fermi surface. At high density, we find metallic spin-unpolarized solutions where electronic charge and spin density form an incommensurate crystal having more crystal sites than electrons. For $r_s\to 0$, our solutions approach pure spin-density waves,  whereas the commensurate Wigner crystal is favored at lower densities, $r_s \gtrsim 3.4$. Decreasing the density, the system undergoes several structural phase transitions with different lattice symmetries. The polarization transition occurs around $r_s \approx 8.5$.

\end{abstract}
\maketitle

The understanding of electrons in solid state and condensed matter has been one of the major challenges since the discovery of quantum mechanics. The simplest model system representing condensed matter is the  homogeneous electron gas (jellium) where  electrons interact with each other and  with a uniform positive charged background  density $3/(4\pi a_B^3r_s^3)$ instead of the nuclei, where $a_B$ is the Bohr radius. For almost one century, jellium has been the central model for qualitative and quantitative studies of electronic correlation\cite{Wigner,Overhauser, QMC2D,QMC3D, exchange,exchange3D, Giulani,momk2D,momk3D,Alavi}.
 
The Hartree-Fock approximation (HF) plays an absolutely fundamental role in tackling many-body electron problems. As the best possible 
description within the independent particle approximation, it provides  both, reference and starting point, for any more sophisticated 
calculations. 
However, even though the HF ground state of jellium has been subject of research all over the years\cite{Overhauser,Needs,Shiwei},
the ground state phase diagram as a function of the density has still not fully been established. 
At low density, 
potential energy largely dominates over the kinetic energy, and the electrons
form the so-called   Wigner crystal (WC), the ground state in the classical limit, whereas in the limit of 
 vanishing $r_s$ the ideal Fermi gas (FG) is approached. 
Overhauser has argued that the FG solution never represents the true HF ground state at any finite density\cite{Overhauser}. 
Only quite recently, indications for a ground state with broken spin symmetry  in the high density region were found in explicit numerical calculations for small and moderate sizes\cite{Shiwei}.  
However, its energy gain compared to FG has not been established in the thermodynamic limit.

Here, we present the Hartree-Fock phase diagram covering relevant crystal structures\cite{Supplementary}.
Generalizing previous approaches to form charge/spin-broken symmetry states\cite{Overhauser,Shiwei},
our study also includes the possibility of incommensurate
crystals of charge/spin-density. In contrast to WC states, the number of maxima of the charge/spin density there differs from the
number of electrons, thus providing broken symmetry states with metallic character\cite{HF-2008,HF-2D}. At high densities, we find that these incommensurate states are favored against FG and WC leading to spin density ground states (SDW). 
Our method allows us to treat large enough systems to obtain results valid in thermodynamic limit, necessary to clearly establish the tiny gain of energy for these states. 
Our study also suggest new candidate ground states for jellium and jellium-like systems\cite{Na}, that should be explored by more accurate many-body approaches\cite{QMC3D,Alavi}.

We consider a system of $N$ electrons in a volume $V$, embedded in an homogeneous background of opposite charge, interacting through the Coulomb potential using periodic boundary conditions.
Hartree-Fock solutions are Slater determinants $\ket{\Psi}=\bigwedge_{\alpha\in S} \ket{\phi_\alpha}$ constituted by a set $S$ of single-particle states $\phi_\alpha$.
In terms of density matrix, the Hartree-Fock solutions can be defined by a 1-body density matrix  $\rho_1$  such that $\Tr \ \rho_1=1$ and $0\leq\rho_1\leq 1/N$. The two-body density matrix $\rho_2$ satisfies:
\begin{align}
\rho_2(\uo,\ut;&\uo',\ut')=\rho_1(\uo;\uo')\rho_1(\ut;\ut')-\rho_1(\uo;\ut')\rho_1(\ut;\uo').
\end{align}
Now we  restrict our study to periodic states. 
Let $\Lambda^*$ be the lattice generated by $\bL_1$, $\bL_2$, $\bL_3$.
Our periodic simulation box
is a parallelepiped of sizes $M\bL_i$, for some integer $M$, and volume $V \sim M^3$,
where we assume $\rho_1(\br+\bL_i,\br'+\bL_i)=\rho_1(\br,\br')$. 
The reciprocal lattice $\Lambda$ is generated by $\bQ_1$, $\bQ_2$, and $\bQ_3$ ($\bL_i \cdot \bQ_j=2\pi \delta_{ij}$) and $\rho_1$ can be written as:
\begin{align}\label{BlockW}
\left(\rho_1\psi\right)(\bk+\bq,\sigma)\!=\!\!\sum_{\bq'\in\Lambda,\sigma'}\!\!\rho_\bk(\bq,\sigma;\bq',\sigma')\psi(\bk+\bq',\sigma')
\end{align}
with $\bk\in\B$, $\bq\in \Lambda$,
where $\B$ is the Brillouin zone of $\Lambda$, and
$\rho_\bk$ are positive matrices satisfying $0\leq\rho_\bk\leq 1/N$.

\begin{figure}
\begin{center}
\includegraphics[scale=0.75]{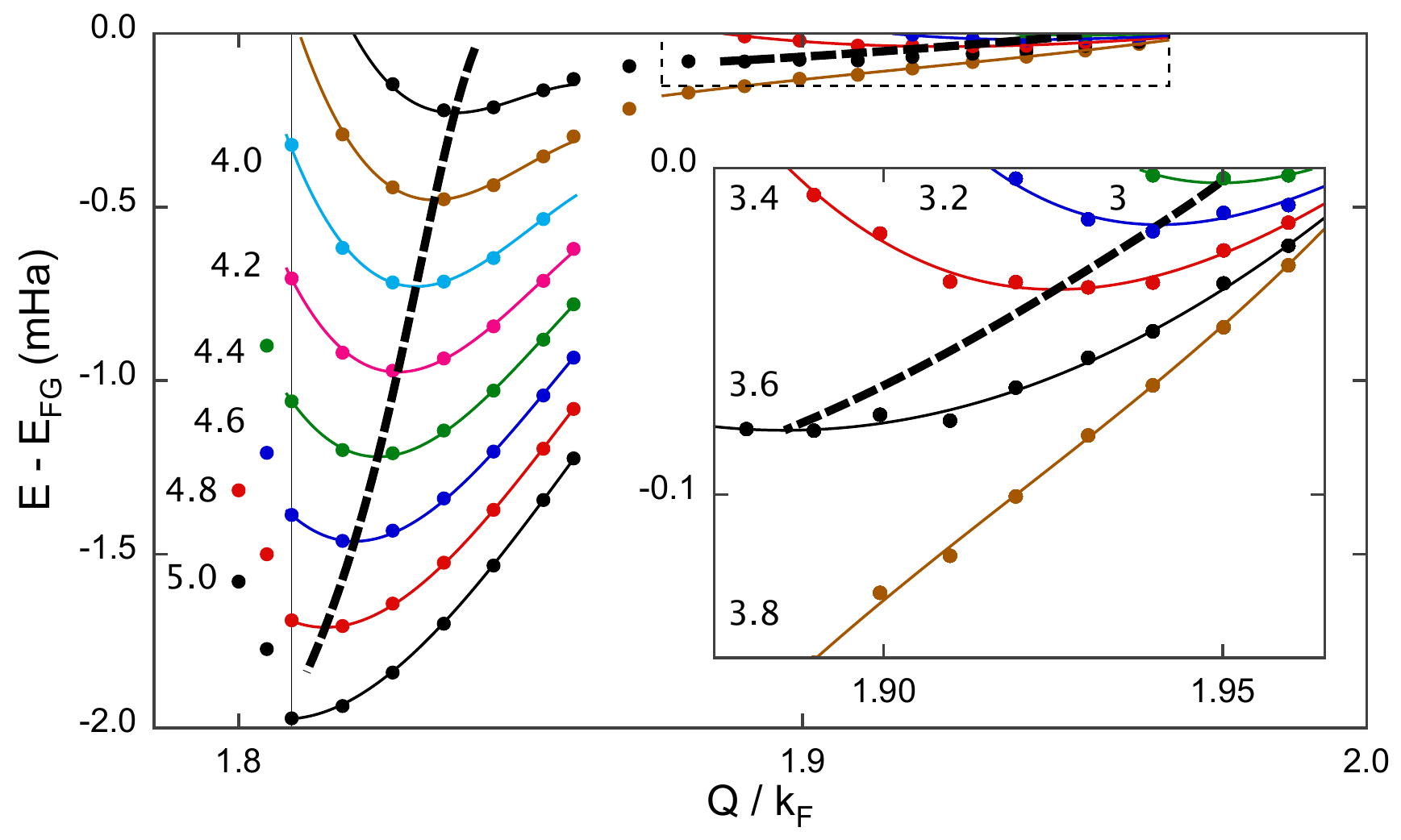}
\caption{
Energy versus the modulation $Q$ at various $r_s$ for the unpolarized gas in the bcc symmetry.
Lines comes from a global polynomial fit  on the numerical results (circles) of order 2 and 3  in $r_s$ and $Q$, respectively. 
$r_s$ is indicated at the start of each curve.
Thick dashed lines go through the local minima.
The leftmost vertical straight line stands for $Q=Q_W$. 
Inset: zoom of the dotted rectangle of the main figure.
\label{FIG-ERSQ-UBCC}
}
\end{center}
\end{figure}

In the following, we concentrate on fully polarized (P) and unpolarized (U) states where $\rho_k$ is restricted to a two component vector,  $\Tr \rho_{k\uparrow}=\Tr \rho_{k\downarrow}$,  but $\rho_{k\uparrow}$ may differ from $\rho_{k\downarrow}$.
We have checked that  the ground state is either U or P except  close to the polarization transition (see Supplementary Materials).
Without any specification, $k_F=(6\pi^2Na_B^3/(n_sV))^{1/3}=\alpha/r_s$, $\alpha^3=9\pi/(2n_s)$, denotes the Fermi wave vector according to the polarization of the corresponding state, with $n_s=2$ for U and $n_s=1$ for P.
For FG solutions, we have
$\rho_{\bk,\sigma}(\bq,\bq')=\delta_{\bq\bq'}\Theta(k_{F}-\|\bk+\bq\|)/N$ 
and the energy per electron is
$E_{FG}=3k_F^2/10-3k_F/(4\pi)=3\alpha^2/(10r_s^2)-3\alpha/(4\pi r_s)$ in Hartree units.

On the other hand, in the Wigner crystal,  each $\rho_\bk$ is $1/N$ times a projector of rank $n_s$.
This case has already been considered with various symmetries in Ref.~\cite{Needs}, but 
their solutions did not lower the energy for $r_s \le 4.4$ and a transition to the FG has been predicted.

Of course, the true ground state solutions are expected to be somewhere between the FG and WC solutions.
Unrestricted HF calculations for small systems\cite{Shiwei} ($N < 10^3$) have indicated the possibility of a spin-density wave
in this region with energy gains of order $10^{-4}\text{Ha}$  with respect to FG.
In fact, at small $r_s$, as the system goes to the FG, the crystalline order remains but the Brillouin zone becomes partially occupied.
In particular, the number of particles per unit cell is not known a priori.
The purpose of this paper is to find these extremal periodic states without extra hypotheses for various lattice symmetries.
In our notation, pure spin density waves (SDW) are U-states verifying  $\rho_{\bk\uparrow}(\bq;\bq')=-\rho_{\bk\downarrow}(\bq;\bq')$ for $\bq\ne\bq'$.

\begin{figure} 
\begin{center}
\includegraphics[scale=1]{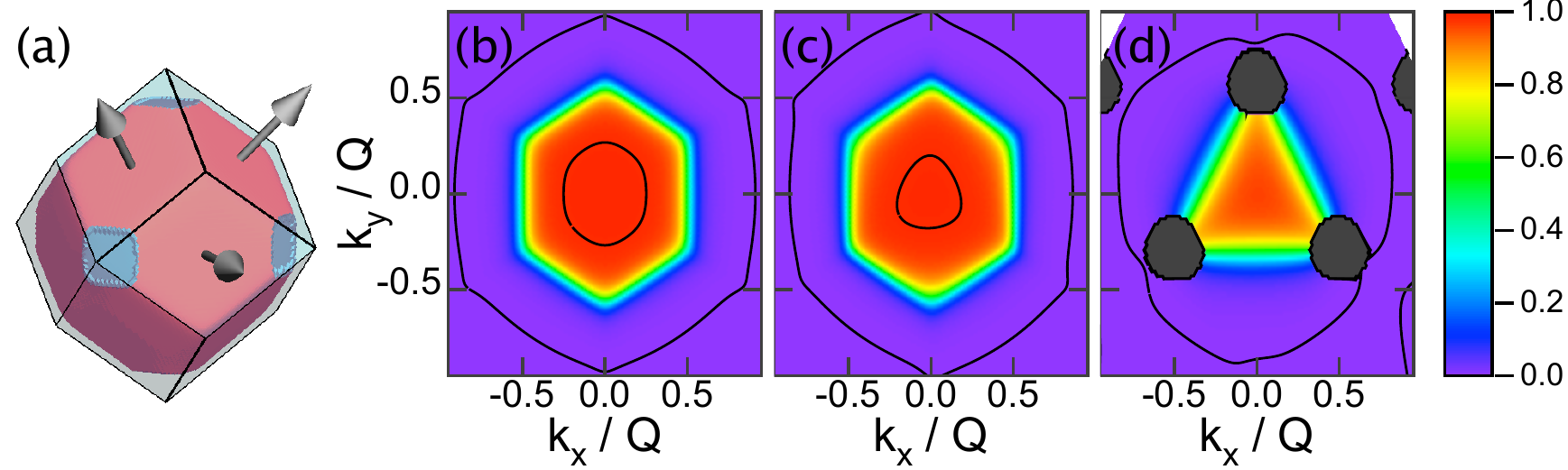}
\includegraphics[scale=1]{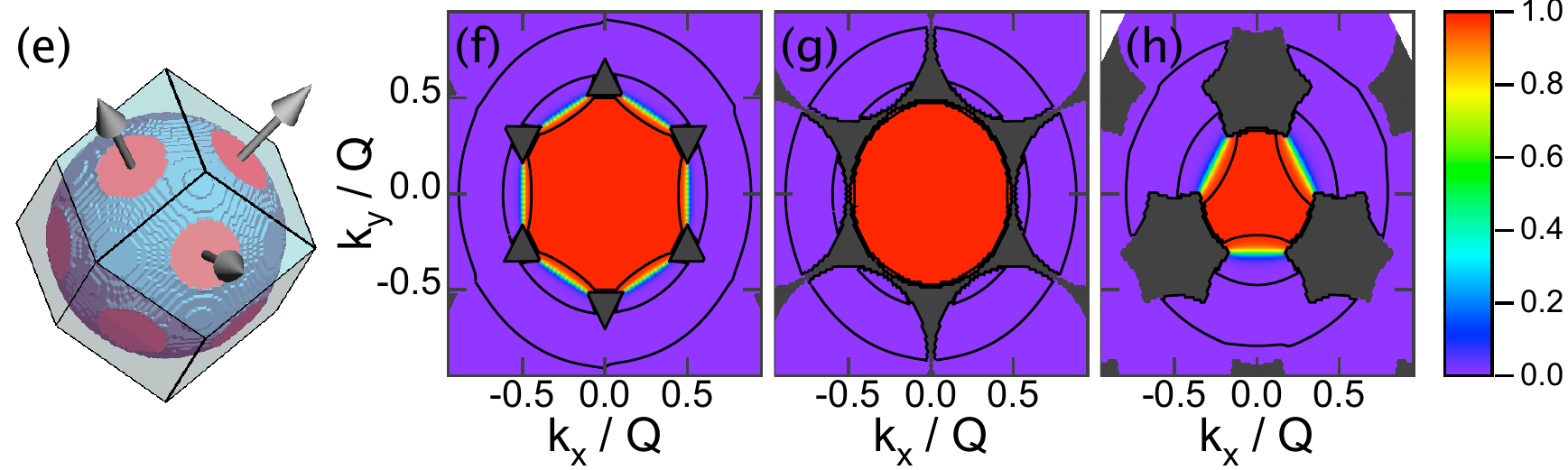}
\caption{
Momentum distribution per spin $n(\bk)$ for incommensurate solutions in bcc symmetry. Figure (a) is an iso-surface at $n(\bk)=0.5$ for $Q/k_F\approx 1.827$, $r_s=4.2$, $M=64$ and $M_\Lambda=19$. The jump of $n(\bk)$ from 0 to a non-zero value are shown in blue. Black arrows stands for reciprocal lattice vectors $(\bQ_1,\bQ_2,\bQ_3)$. On (b), (c), and (d) : cut of $n(\bk)$  in the plane $(\bQ_1,\bQ_2)$ at $k_3=0$, $Q/4$, and $Q/2$ respectively.
Black areas correspond to $n(\bk)=0$. 
Contour levels are at 0.1, $10^{-3}$ and $10^{-5}$.
(e-h): same as (a-d) for $Q/k_F\approx 1.940$ and $r_s=3.2$.
\label{FIG-RHOK-CUBIC}
}
\end{center}
\end{figure}

Thus, we search for a lattice $\Lambda$ and a density matrix $\rho_\bk$  such that the number of particles per unit cell is near $n_s$ (or some multiple of $n_s$ for non-Bravais lattices (nBL)). Notice that  for extremal states, the eigenvalues of $\rho_\bk$ must be exactly $0$ or $1/N$. 
The number of strictly positive eigenvalues
is not known a priori, but is expected to fall between 0 and $2n_s$ (or some multiple of $2n_s$ for  nBL).

We truncate the number of vectors of the sub-lattice $\Lambda$, including only the first $\NL$ bands: $\rho_\bk$ is a square matrix of order $n_s\NL$.
The condition $0\leq\rho_\bk\leq1/N$, is difficult to fulfill. 
So we choose the representation: 
\begin{align}
\label{EQ-rhokUDU}
\rho_\bk=\sum_i D_{\bk,i}\ket{u_{\bk,i}}\bra{u_{\bk,i}}
\end{align}
where $\braket{u_{\bk,i}}{u_{\bk,j}}=\delta_{ij}$ and $0\leq D_{\bk,i}\leq1/N$.
Since the number of strictly positive $D_{\bk,i}$ is between 0 and $2n_s$, we can restrict the summation in Eq. (\ref{EQ-rhokUDU}) over $2n_s$ terms instead of $n_s\NL$. The number of unknowns is thus of order $2n_s\NL$ times the number of vectors of $\B$. This is why we can deal with large number of particles\cite{DIFF-FROM_2D}.

The minimization consists 
in the following steps. At first we choose $D_{\bk,i}$ and $\ket{u_{\bk,i}}$ to start with. Then, for $D_{\bk,i}$ fixed, 
we find the best $\ket{u_{\bk,i}}$ with a quadratic descent method\cite{HF-2008}. The next step is to try to improve $D_{\bk,i}$ given the gradient of the energy
with respect to $D_{\bk,i}$ and the linear constraints, $0\le D_{\bk,i}\le1/N$ and  $\sum_{\bk,i} D_{\bk,i}=1$.
We thus obtain a new set  $D_{\bk,i}^{(\rm new)}$ (either 0 or $1/N$),
and we change $D_{\bk,i}$ into  
$(1-\varepsilon)D_{\bk,i}+\varepsilon D_{\bk,i}^{(\rm new)}$ (with a small enough $\varepsilon$ to ensure 
that $\ket{u_{\bk,i}}$ follows  $D_{\bk,i}$ {\sl adiabatically}) and we restart the minimization with respect to $\ket{u_{\bk,i}}$.
The process stops as soon as  $D_{\bk,i}^{(\rm new)}=D_{\bk,i}$. In this case almost every $D_{\bk,i}$ are
$0$ or $1/N$ and the gradient is negative or positive accordingly. 

\begin{figure}
\begin{center}
\includegraphics[scale=1]{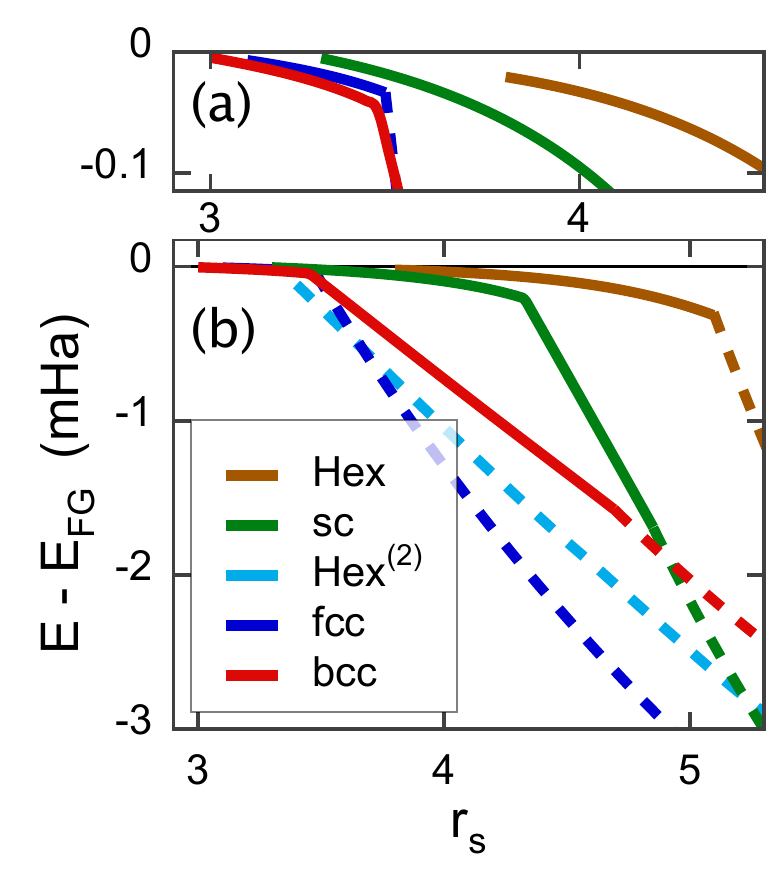}
\includegraphics[scale=1]{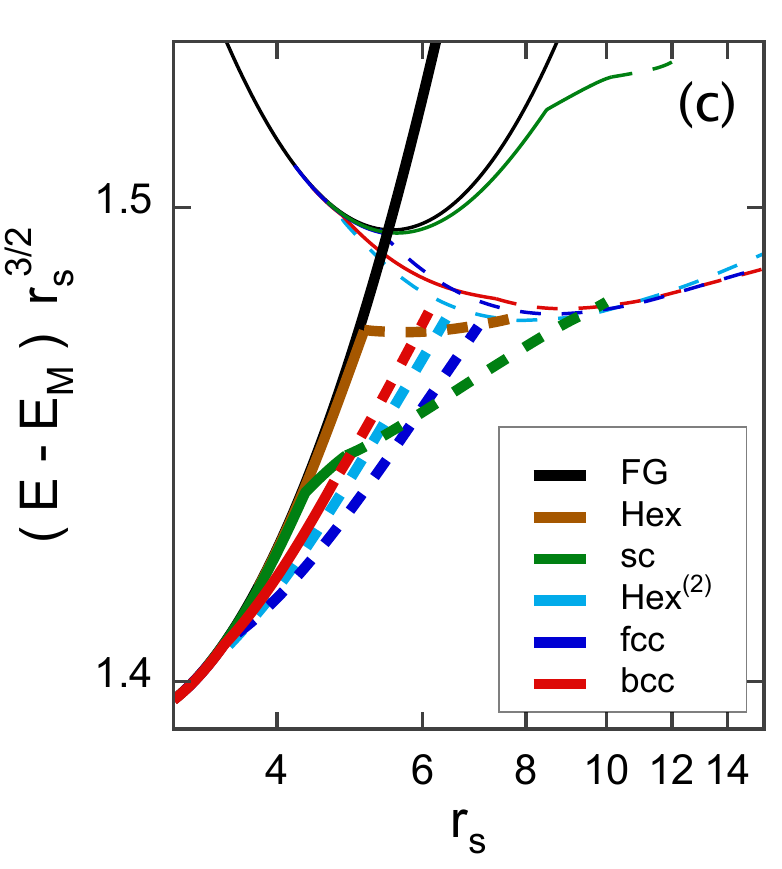}
\caption{Phase diagram of the electron gas: unpolarized (a) and (b), polarized and unpolarized (c).
Energies are in milli-Hartrees for (a) and (b), and in Hartrees for (c), where $E_M=0.89593/r_s$ is the Madelung energy of a polarized-bcc Wigner crystal.
Full lines stand for incommensurate regime ($Q>Q_W$) and dashed lines for the Wigner crystal ($Q=Q_W$).
Colors refer to the lattice (see Table \ref{TAB-LATTICE-SPINCHARGE}). 
(a): zoom of (b) around $E-E_{FG}=0$.
(c): thin lines stand for the polarized gas (upper curves) and thick lines for the unpolarized gas\cite{Supplementary}.
\label{FIG-PDIAGRAM}
}
\end{center}
\end{figure}

\begin{table}
\begin{center}
\begin{tabular}{c|c|c|c|c|c}
\multicolumn{6}{c}{Unpolarized} \\
\hline
Symmetry   & IC-bcc& \HT(hcp$^*$)& fcc & sc  & Hex \\
\hline
$n_m$ & 2 & 4 & 2 & 2 & 2\\
Charge & IC-bcc$^{(2)}$ & \HH & sc  &bcc& hcp \\
$T$ & ($\frac12,\frac14,0$) & $T_2$& ($\frac12,0,0$)& ($\frac12,\frac12,\frac12$)  & $T_1$ \\
$r_s$&$3-3.4$ &  $3.4-3.7$ & $3.7-5.9$ & $5.9-9.3$ &
\end{tabular}
$\qquad\qquad$
\begin{tabular}{c|c|c|c|c}
 \multicolumn{5}{c}{Polarized} \\
\hline
Symmetry  &sc& \HT(hcp) & fcc & bcc \\
\hline
$n_m$& 1&2  & 1 & 1 \\
$Q_W/k_F$ & $1.61$ & $0.88$ & $1.76$ & $1.81$\\
$r_s$ && $9.3-10.3$&$10.3-13$&$13-$ 
\end{tabular}
\caption{
The best overall ground states depending  on $r_s$ (last line).
(IC) stands for incommensurate crystal, otherwise it is WC.
$n_m$ is the number of maxima of charge density per unit cell\cite{NM-WI}.
$T$ is the shift between the spin up and down lattices, in the conventional (cartesian) cell basis for sc, fcc and bcc lattices, and in the primitive cell basis for Hex and \HT (number in parenthesis is the number of sites per cell). 
$T_1=(\frac{1}{3},\frac{2}{3},\frac{1}{2})$, $T_2=(\frac{1}{2},0,\frac{1}{4})$. 
The star means close to.
The last line of each table indicates the range of $r_s$ where each phase may be in the ground state.
\label{TAB-LATTICE-SPINCHARGE} 
}
\end{center}
\end{table}

The parameters are $r_s$ (for the density), the lattice symmetry, the (smallest) modulus $Q$ of the generators of $\Lambda$, the number $M^3$ of points in the Brillouin zone $\B$ and the number $M_\Lambda$ of plane waves per single-particle state.

A priori, we look for lattices with the lowest Madelung energies as they will lead to the more stable states at low densities. However, as the density increases, other lattices may become more favorable. 
Investigated lattices are: simple cubic (sc), face-centered cubic (fcc), body centered cubic (bcc) and hexagonal (Hex) (see Table \ref{TAB-LATTICE-SPINCHARGE}). 

For WC phases, $Q=Q_W$, whereas $Q \ne Q_W$ characterizes incommensurate crystals, and $Q \ge 2k_F$ leads  to the FG solution with isotropic Fermi surface at $k_F$. 
Increasing $M_\Lambda$ increases the basis resulting in a lower energy due to the variational principle.
Our discretization of the Brillouin zone ranges from $M=32$ up to $M=128$ which corresponds to effective system sizes with number of electrons ($\sim M^3$) much larger than those of Ref.~\cite{Shiwei}. 

\begin{figure}
\begin{center}
\includegraphics[scale=0.5]{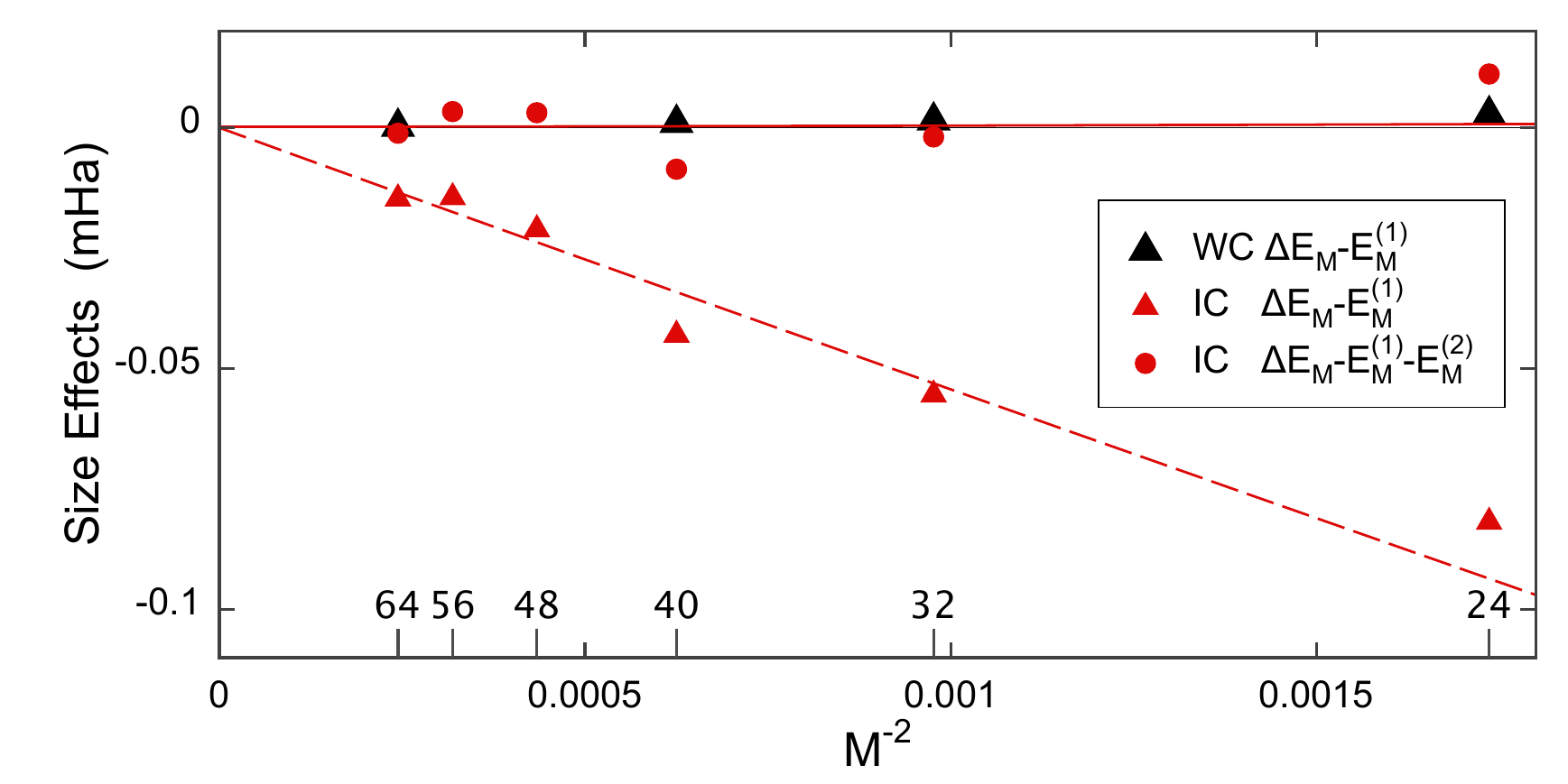}
\caption{Size effects on the potential energy for the U-bcc lattice, Eqs.~(\ref{EQ-EM1}-\ref{EQ-EM2}).
$E_M^{(i)}$ is the correction  of order $1/M^{i}$.
In red and black are the data for IC $[r_s=3.4, Q/k_F=1.92]$ and WC $[r_s=5]$, respectively.
Red circles include second order correction for IC ($E_M^{(2)}=0$ for WC).
The red dashed (resp. full) line is a fit of the form $E_M^{(2)}+b/M^3$ (resp. $b/M^3$).
Numbers at the bottom indicate $M$.
\label{FIG-convm}
}
\end{center}
\end{figure}

Finite size effects are important in fermionic Coulomb systems\cite{Simone} and, contrary to $M_\Lambda$, there is no variational principle.
As the memory size increases  $\propto M_\Lambda^2M^3$, pure numerical extrapolation to the thermodynamic limit ($M \to \infty$) is difficult.
Therefore, to accelerate convergence, we have included finite size corrections:
\begin{align}
\label{EQ-EM1}
	\Delta E_M \equiv E_M-E_\infty= E_M^{(1)} + E_M^{(2)} + E_{\rm NA}
\end{align}
where $E_M^{(1)} \sim M^{-1}$ is the Madelung energy,
$E_M^{(2)}$ is an analytical potential energy error of order $M^{-2}$, and $E_{\rm NA}$ contains the non-analytical enery corrections of order $M^{-3}$.
From the FG-potential energy, $E_M^{(2)}$ can be estimated as:
\begin{align}
\label{EQ-EM2}
	E_M^{(2)}=-\left(\frac{ \gamma}{\pi M} \left.\frac{S(\bk)}{\|\bk\|} \right|_{k\rightarrow0} \right) \times E_M^{(1)}
\end{align}
where $\gamma^3$ is the volume of $\B$, and $S(\bk)$ is the structure factor (for FG, $\lim_{k\rightarrow 0}  S(\bk) / \|\bk\| = 3 / (4k_F)$).
Notice that $E_M^{(2)}$ is maximum for FG, decreases with $Q$ for incommensurate solutions, and vanishes for WC. 
As can be seen in 
Fig.\ref{FIG-convm}, removing $E_M^{(2)}$ greatly improves the thermodynamic limit extrapolation. 
However, the remaining non-analytical  contributions of order $10^{-6}$Ha become comparable to the energy gain at high densities, and prevent a precise determination of the ground state for $r_s \lesssim 3$.
Extending the analytical calculations of Ref.~\cite{HF-2008-2} from two to three dimensions, one can prove that the incommensurate
phases are always energetically favorable for $r_s \to 0$.

The accuracy of our results is essentially controlled at large $r_s$ by $M_\Lambda$ ($\rho_\bk$ smooth but extended) and at small $r_s$ by $M$ ($\rho_\bk$ rapidly varying around $k_F$).
Fig.(\ref{FIG-ERSQ-UBCC}) shows the energy differences  $\Delta E=E-E_{FG}$ versus the modulation $Q$ at various $r_s$ for the U-bcc symmetry.
At large $r_s$, the minimum is found for $Q=Q_W$. At smaller $r_s$, a minimum is eventually reached for $Q>Q_W$.
As seen in Fig. (\ref{FIG-ERSQ-UBCC}), two local minima may occur.
Furthermore, at the local minima, the best solutions are always found with only one band (i.e., for each $\bk\in\B$ in Eq.~(\ref{EQ-rhokUDU}), at most one $D_{\bk,i}$ is non-zero).

The momentum distribution $n(\bk)$ shows how the incommensurate states interpolate between the Wigner phase and the Fermi gas as $r_s$ decreases ($Q$ going from $Q_W$ to $2k_F$).
For WC, the first band in $\B$ is fully occupied and $n(\bk)$ is everywhere continuous. 
At $Q\gtrsim Q_W$, the first band becomes partially filled and unoccupied volumes appear around the corners of $\B$ (see Fig \ref{FIG-RHOK-CUBIC}-a,d), leading to a possible minimum at some $\QM$. 
Thus $n(\bk)$ is discontinuous at the surface of these volumes but stay continuous in other directions.
As $Q$ increases, the volumes connect allowing an eventual second minimum at $Q\lesssim 2k_F$ (see Fig \ref{FIG-RHOK-CUBIC}-e-g).
At $Q=2k_F$, the Fermi sphere is completed, where $n(\bk)$ is discontinuous at the Fermi surface.

\begin{figure}
\begin{center}
\includegraphics[scale=0.75]{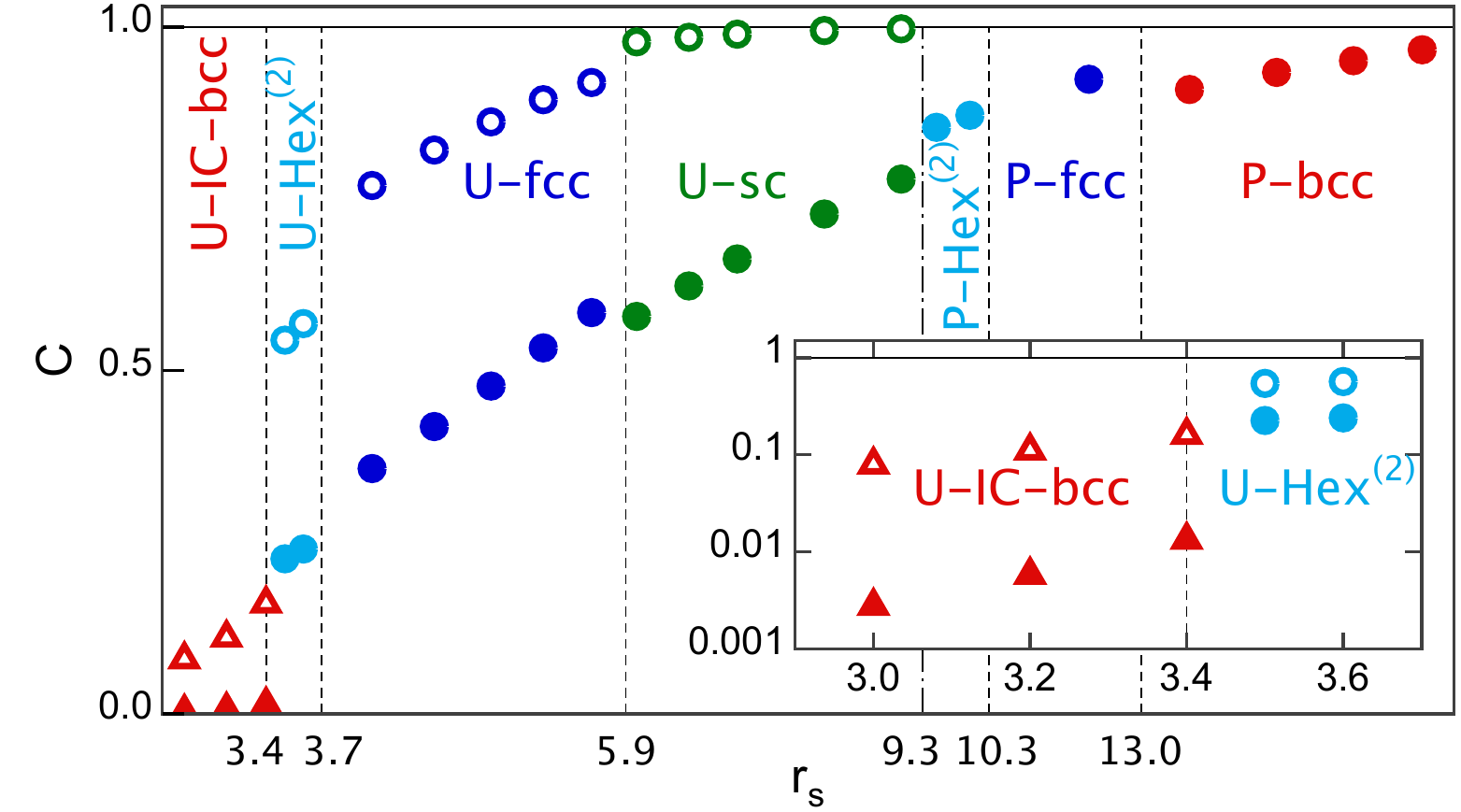}
\caption{Contrast $C$ of $n(\br)$ versus $r_s$ for the ground states.
Full symbols: charge density.
Open symbols (for unpolarized gas): spin up (or down) density.
Circle: Wigner Crystal.
Triangle: Incommensurate phase (IC).
Inset: zoom of the low-$r_s$ part.
Vertical  lines separate the different unpolarized (U) and polarized (P) phases.
In the incommensurate phase, charge modulations are drastically reduced at higher densities
and the solutions approach a  pure SDW ground state.
\label{FIG-nx}
} 
\end{center}
\end{figure}

By construction, the real-space density, $n(\br)$, has the crystalline symmetry of the lattice.
In the incommensurate phase (IC), the number of maxima is greater than the number of electrons and depends on $Q$.
At large $r_s$, the numbers coincide, this is the Wigner crystal.
We define a contrast by $C=(n^{\max}-n^{\min})/(n^{\max}+n^{\min})$ which goes from 0 to 1 as $r_s$ goes from 0 to infinity, where $n^{\max}$ and $n^{\min}$ are the maxima and minima of $n(\br)$.
For the unpolarized gas, we define the contrast for each spin-species and for the total charge.
As shown in Fig.~(\ref{FIG-nx}), the contrast decreases rapidly as $r_s$ goes to zero; it is expected\cite{HF-2008-2} to be a non-analytic function at $r_s=0$.
Notice that the charge density modulation is much smaller than each spin modulation in the unpolarized gas
demonstrating the SDW character of the ground state.

The final phase diagram of the unpolarized and polarized gas is reported in Fig.\ref{FIG-PDIAGRAM} and Table \ref{TAB-LATTICE-SPINCHARGE}.
At high density, the incommensurate states have SDW character with modulations $Q>Q_W$ which increase at smaller $r_s$ towards $Q=2k_F$. As only states close to the Fermi surface are relevant in this region, energy gains compared to FG become very tiny. 
Our resolution in $k$-space is insufficient to determine the precise modulations for $r_s < 3$, which introduce small anisotropies in the Fermi surface for $Q < 2k_F$.
Nevertheless, our calculations explicitly confirm the instability of the FG towards SDW\cite{Overhauser,Giulani}
and indicate that the spin modulation continuously approaches $Q=2 k_F$ with isotropic Fermi surface for $r_s \to 0$. 

To conclude, we have established the true ground state phases of jellium within Hartree-Fock over a broad density region. 
In particular, we have shown that the Overhauser instability\cite{Overhauser} of the FG results in a new ground state in the thermodynamic limit, characterized by an incommensurate crystal structure for the spin and charge density. However, it is known that Hartree Fock tends to favor crystalline phases, as the gain in correlation energy is typically higher in the isotropic FG phase than in the WC\cite{QMC3D}. Therefore, the transition to the WC is quantitatively incorrect within HF and shifted towards considerable higher values of $r_s$ in the true ground state phase diagram. Whereas correlations certainly stabilize the FG at small $r_s$,
correlations should favor incommensurate phases compared to WC for the same reason, so that 
incommensurate states should actually occur at densities close to crystallization. 
We hope that  
future QMC calculations will be able to establish this new phase beyond HF.

\newpage

\section{Supplementary Materials: Total spin of the HF-states}
Without imposing the polarization, each $\rho_k$ may be written as a matrix
\begin{align}
	\rho_k(q,q')=\left(
		\begin{array}{cc}
			\rho_{k,\upuparrows}(q,q') & \rho_{k,\updownarrows} (q,q')\\
			\rho_{k,\updownarrows}^*(q,q') & \rho_{k,\downdownarrows} (q,q')
		\end{array}
	\right)
\end{align}
The total spin square reads $S^2=\sum_{\alpha=1}^3(\sum_i s_i^\alpha)^2$
where $s_i^\alpha$ are the components of the spin of particle $i$,
and we have
\begin{align}
	s^2&=\frac{\left<S^2\right>}{N^2}=\frac1{N^2}\Tr S^2\rho_N^{}
		=\frac{N-1}N\sum_{a=1}^3\Tr\rho_2s^\alpha\otimes s^\alpha+\frac3{4N}
\end{align}
Inserting Eq. (1) of the main text, we get in the thermodynamic limit (the exchange term vanishes):
\begin{align}
	s^2&\begin{array}{c}\\\longrightarrow\\{{}^{N\to\infty}}\\\end{array}
		\sum_{\alpha=1}^3\left(\Tr\rho_1s^\alpha\right)^2
\label{EQ-def-s2}
		=\frac{(\Tr\rho_{\upuparrows}-\Tr\rho_{\downdownarrows})^2}4+|\Tr\rho_{\updownarrows}|^2
\end{align}
The polarized case (P), $s^2=1/4$, can be obtained with various total spin orientation. One can show that after a rotation, one has 
$\Tr\rho_{\upuparrows}=1$ and $\Tr\rho_{\downdownarrows}=\Tr\rho_{\updownarrows}=0$.
The unpolarized case, $s^2=0$ is obtained with $\Tr\rho_{\upuparrows}=\Tr\rho_{\downdownarrows}$ (same number of up and down spins) and $\Tr\rho_{\updownarrows}=0$.
Many states are possible where $\rho_{\updownarrows}\ne0$ .
In the paper, unpolarized states  (U) means $\rho_{\updownarrows}=0$.
Here, we call restricted the U and P states and unrestricted the general model (Eq.1).

\begin{figure}[h]
\begin{center}
\begin{minipage}[t]{0.475\textwidth}
\includegraphics[scale=0.45]{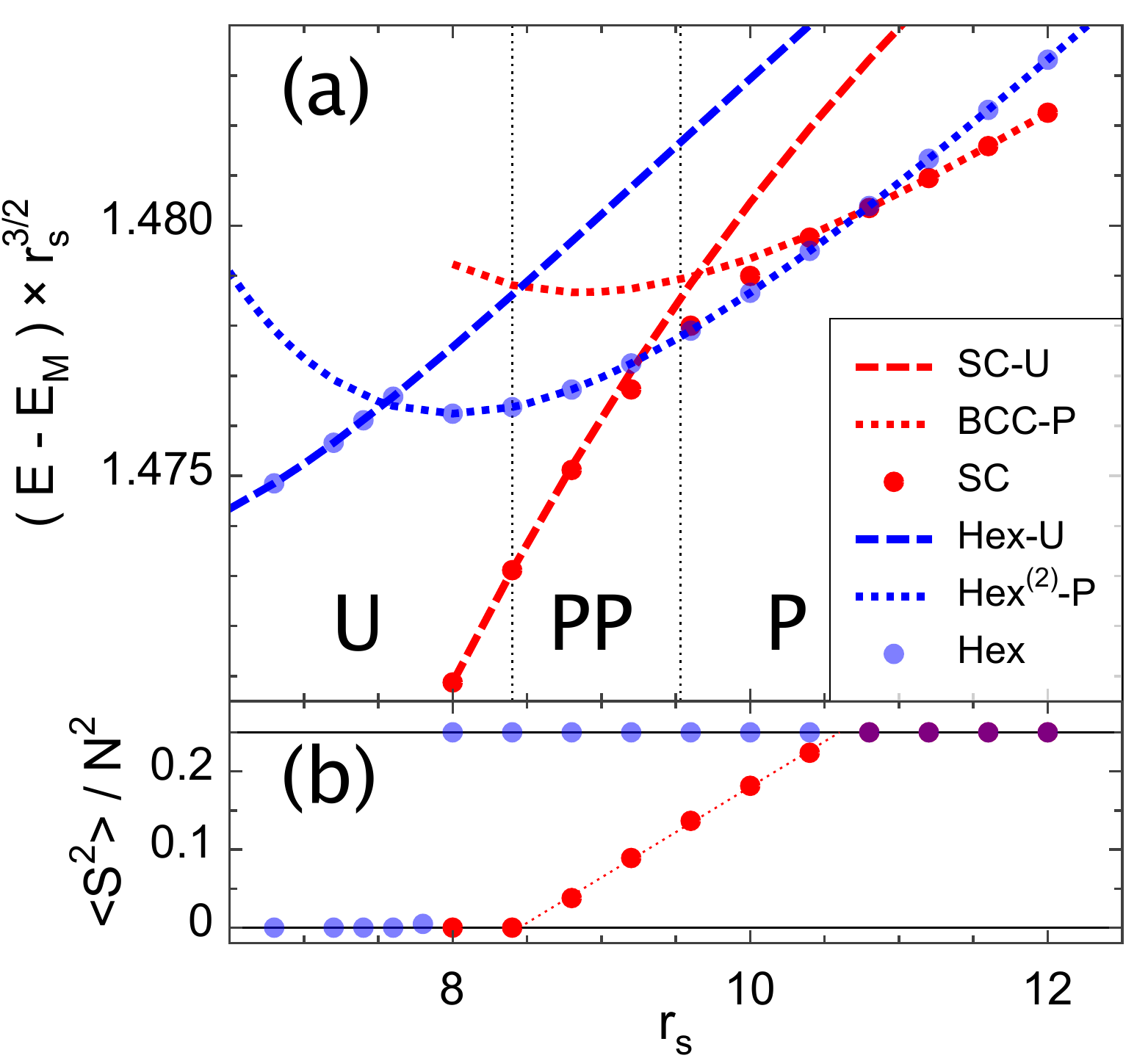}\ \ \ \ \ \ 
\end{minipage}
\begin{minipage}[t]{0.475\textwidth}
\vspace*{-170px}
\begin{tabular}{r|c|c|c|c}
\multicolumn{2}{c|}{} &\multicolumn{3}{c}{$E$ (mHa)}\\ \cline{3-5}
$r_s$ & $\left<S^2\right>/N^2$&unrestricted& U	&P\\
\hline
8.0	& 0.0000	& -46.987	& -46.987	& -46.618\\
8.4	& 0.0000	& -46.150	& -46.150	& -45.916\\
8.8	& 0.0379	& -45.303	& -45.301	& -45.167\\
9.2	& 0.0891	& -44.464	& -44.451	& -44.392\\
9.6	& 0.1367	& -43.636	& -43.608	& -43.603\\
10.0	& 0.1815	& -42.823	& -42.777	& -42.812\\
10.4	& 0.2238	& -42.026	& -41.961	& -42.025\\
10.8	& 0.2500	& -41.247	& -41.164	& -41.247\\
11.2	& 0.2500	& -40.483	& -40.386	& -40.483\\
11.6	& 0.2500	& -39.734	& -39.629	& -39.734\\
12.0	& 0.2500	& -39.003	& -38.893	& -39.003\\
\end{tabular}
\end{minipage}
\caption{
Results with $M=16$ for SC symmetry in red ($M_\Lambda=57$), 
Hex symmetry in blue ($M_\Lambda=35$). 
Dashed and dotted lines and circles stand for U, P, and the unrestricted model, respectively.
(a): Energy versus $r_s$, $E_M=0.89593/r_s$ is the P-BCC Madelung energy.
Vertical dotted lines separate the domains where the overall ground state is unpolarized (U), polarized (P) or the partially polarized one (PP).
(b): $\left<S^2\right>/N^2$ (see Eq.\ref{EQ-def-s2}) versus $r_s$. The red dotted line is a linear fit through the partially polarized points.
Energies per particle and $\left<S^2\right>/N^2$ for SC symmetries are summarized in the table.
\label{FIG-1}
}
\end{center}
\end{figure}

The phase diagram, Fig.~3 of the main article has been obtained with restricted (U and P) states.
Since the U-P transition happens around the density region $8<r_s<12$, we focus on it by performing unrestricted calculations.
With SC symmetry (leading to solutions with BCC charge symmetry), we can follow the transition from U-SC to P-BCC.
Fig.1-a shows the energies versus $r_s$ all models.
Fig.1-b shows the variations of $s^2$ where a crossover is seen for  $8.5<r_s<10.6$, the domain where the unrestricted model leads to lower energy states than the restricted ones.
The other candidate for the absolute ground state is with Hex symmetry where the ground state goes for U-Hex directly to P-Hex$^{(2)}$ (see Fig.1a-b).
The comparison of all these curves shows a transition for U-SC to P-Hex$^{(2)}$ within a small region ($8.4 <r_s<9.5$) where the polarization of the SC phase increases.
More work is under investigation to understand this partially polarized phase. 

For the remaining U-BCC, U-FCC, U-Hex$^{(2)}$ symmetries, we have checked on small system ($M=8$) that the unrestricted results do not improve the energies.
\newpage
\section{Supplementary Materials: Ground state energies versus $r_s$}

\begin{table*}[h]
\begin{center}
\begin{tabular}{c|c|c|c|c}
\multicolumn{5}{c}{Unpolarized} \\
$r_s$ & $E$(mHa) & $\Delta E$(mHa)& $Q/k_F$ & sym. \\
\hline
3.0		& -29.954	& -0.005	& 1.9495	& IC-bcc	\\ 
3.1		& -32.826	& -0.010	& 1.9456	& IC-bcc	\\ 
3.2		& -35.289	& -0.017	& 1.9406	& IC-bcc	\\ 
3.3		& -37.399	& -0.026	& 1.9341	& IC-bcc	\\ 
3.4		& -39.287	& -0.117	& 0.8794	& Hex$^{(2)}$	\\ 
3.5		& -40.923	& -0.272	& 0.8794	& Hex$^{(2)}$	\\ 
3.6		& -42.437	& -0.427	& 0.8794	& Hex$^{(2)}$	\\ 
3.7		& -43.727	& -0.611	& 1.7589	& fcc	\\ 
3.8		& -44.899	& -0.849	& 1.7589	& fcc	\\ 
4.0		& -46.775	& -1.293	& 1.7589	& fcc	\\ 
4.2		& -48.157	& -1.709	& 1.7589	& fcc	\\ 
4.4		& -49.151	& -2.096	& 1.7589	& fcc	\\ 
4.6		& -49.841	& -2.459	& 1.7589	& fcc	\\ 
4.8		& -50.292	& -2.799	& 1.7589	& fcc	\\ 
5.0		& -50.554	& -3.119	& 1.7589	& fcc	\\ 
5.2		& -50.665	& -3.420	& 1.7589	& fcc	\\ 
5.4		& -50.656	& -3.703	& 1.7589	& fcc	\\ 
5.6		& -50.551	& -3.970	& 1.7589	& fcc	\\ 
\end{tabular}
$\quad$
\begin{tabular}{c|c|c|c|c}
\multicolumn{5}{c}{Unpolarized} \\
$r_s$ & $E$(mHa) & $\Delta E$(mHa) & $Q/k_F$ & sym. \\
\hline
5.8		& -50.368	& -4.221	& 1.7589	& fcc	\\ 
6.0		& -50.192	& -4.524	& 1.6120	& sc	\\ 
6.2		& -50.031	& -4.878	& 1.6120	& sc	\\ 
6.4		& -49.813	& -5.201	& 1.6120	& sc	\\ 
6.6		& -49.550	& -5.497	& 1.6120	& sc	\\ 
6.8		& -49.249	& -5.768	& 1.6120	& sc	\\ 
7.0		& -48.919	& -6.017	& 1.6120	& sc	\\ 
7.2		& -48.566	& -6.247	& 1.6120	& sc	\\ 
7.4		& -48.193	& -6.457	& 1.6120	& sc	\\ 
7.6		& -47.804	& -6.649	& 1.6120	& sc	\\ 
7.8		& -47.403	& -6.825	& 1.6120	& sc	\\ 
8.0		& -46.992	& -6.987	& 1.6120	& sc	\\ 
8.2		& -46.576	& -7.135	& 1.6120	& sc	\\ 
8.4		& -46.155	& -7.271	& 1.6120	& sc	\\ 
8.6		& -45.731	& -7.396	& 1.6120	& sc	\\ 
8.8		& -45.307	& -7.511	& 1.6120	& sc	\\ 
9.0		& -44.883	& -7.617	& 1.6120	& sc	\\ 
9.2		& -44.461	& -7.716	& 1.6120	& sc	\\ 
\end{tabular}
$\quad$
\begin{tabular}{r|c|c|c|c}
\multicolumn{5}{c}{Polarized} \\
$r_s$ & $E$(mHa)& $\Delta E$(mHa)& $Q/k_F$ & sym. \\
\hline
9.4		& -44.050	& -7.814	& 0.8794	& Hex$^{(2)}$	\\ 
9.6		& -43.647	& -7.911	& 0.8794	& Hex$^{(2)}$	\\ 
9.8		& -43.245	& -7.998	& 0.8794	& Hex$^{(2)}$	\\ 
10.0		& -42.844	& -8.077	& 0.8794	& Hex$^{(2)}$	\\ 
10.2		& -42.444	& -8.146	& 0.8794	& Hex$^{(2)}$	\\ 
10.4		& -42.047	& -2.759	& 1.7589	& fcc	\\ 
10.7		& -41.461	& -2.833	& 1.7589	& fcc	\\ 
11.0		& -40.883	& -2.901	& 1.7589	& fcc	\\ 
11.5		& -39.936	& -3.003	& 1.7589	& fcc	\\ 
12.0		& -39.015	& -3.092	& 1.7589	& fcc	\\ 
12.5		& -38.126	& -3.171	& 1.7589	& fcc	\\ 
13.0		& -37.267	& -3.242	& 1.7589	& fcc	\\ 
13.5		& -36.441	& -3.306	& 1.8094	& bcc	\\ 
14.0		& -35.645	& -3.362	& 1.8094	& bcc	\\ 
14.5		& -34.880	& -3.412	& 1.8094	& bcc	\\ 
15.0		& -34.142	& -3.455	& 1.8094	& bcc	\\ 
15.5		& -33.432	& -3.491	& 1.8094	& bcc	\\ 
16.0		& -32.748	& -3.522	& 1.8094	& bcc	\\ 
\end{tabular}
\caption{Ground state energies, E, at the thermodynamic limit, in milli-Hartree units (mHa)
and energy gain, $\Delta E=E-E_{FG}$, compared to the Fermi gas solution (precision: $\sim5$ on the last digit)
$Q$ is the modulus of the generator of $\Lambda$ (see Eq.\ (2) of the main article).
sym. means lattice symmetry.
IC indicates incommensurate crystalline order.
\label{HFenergies}
}
\end{center}
\end{table*}

Following are details on the size effects as defined in Eqs. 4-5 of the main text.
We recall:
\begin{align}
E_M^{(1)}&=-\frac{\alpha\gamma Q C_\Lambda}{4\pi r_s M} \\
E_M^{(2)}&=-\frac{\gamma Q S_0}{\pi M} \  E_M^{(1)}
\end{align}
where $\alpha^3=9\pi/(2n_s)$, $\gamma^3=\det(M_Q)$ and
$C_\Lambda$, the Madelung constant, are given in the Tab.~\ref{TAB-LATTICE}, and $S_0 =\lim_{\bk\rightarrow0} S(\bk)/\|\bk\|$. $S_0$ is given in Tab.~\ref{S0} for the incommensurate phase (IC-bcc) and $S_0=0$ in the Wigner phase ($Q=Q_W$).
\begin{table}[htbp]
\begin{center}
\renewcommand{\arraystretch}{1.5}
\begin{tabular}{c||c|c|c|cc}
lattice
& sc
& bcc
& fcc
& Hex& Hex$^{(2)}$
\\
\hline
$n_c$&1&1&1&1&2\\
$M_Q$
& $\left(\begin{array}{rrr}
	1 & 0 & 0 \\
	0 & 1 & 0 \\
	0 & 0 & 1
	\end{array}\right)$
& \scalebox{1.4}{
$\frac{1}{\sqrt2}$
}
$\left(\begin{array}{rrr}
	0 & 1 & 1 \\
	1 & 0 & 1 \\
	1 &1 & 0
	\end{array}\right)$
& \scalebox{1.4}{$\frac{1}{\sqrt3}$}
	$\left(\begin{array}{rrr}
	-1 & 1 & 1 \\
	1 & -1 & 1 \\
	1 & 1 & -1
	\end{array}\right)$
& \multicolumn{2}{c}{\scalebox{1.4}{$\frac{4\sqrt2}{3}$}
	$\left(\begin{array}{ccc}
	1 & 1/2 & 0 \\
	0 & \sqrt{3}/2 & 0 \\
	0 & 0 & 3/(4\sqrt{2})
	\end{array}\right)$}
\\
$\gamma^3$&1&$\frac1{\sqrt2}$&$\frac4{3\sqrt3}$&\multicolumn{2}{c}{$\frac{16}{3\sqrt3}$}\\
$Q_W/k_F$ 
& $1.611991954016$
& $1.809399790564$
& $1.758882522024$
& $1.108026556895$
& $0.879441261012$
\\
$C_\Lambda$
& -2.837297479481
& -2.888461503054
& -2.888282119020
& \multicolumn{2}{c}{-2.512880623796}
\end{tabular}
\caption{Lattice definitions and properties. 
$n_c$ is the number of electrons per primitive cell.
$M_Q=(\bQ_1/Q, \bQ_2/Q, \bQ_3/Q)$. 
For the hexagonal case, $Q=\|\bQ_3\|$. 
$Q_W$ is given by: $Q_W/k_F=\gamma(4\pi/(3n_c))^{1/3}$, $k_F=\alpha/r_s (a.u.)$.
}
\label{TAB-LATTICE}
\end{center}
\end{table}

\begin{table}
\begin{tabular}{c||c|c|c|c}
$r_s$ & 3.0 & 3.1 & 3.2 & 3.3 \\
\hline
$S_0 k_F$ & 0.596(5) & 0.567(5) & 0.540(5) & 0.512(5)
\end{tabular}
\caption{
$S_0=\lim_{\bk\rightarrow0} S(\bk)/\|\bk\|$ versus $r_s$ for IC-U-bcc. 
\label{S0}
}
\end{table}


\begin{thebibliography}{99}
\bibitem{DIFF-FROM_2D}
This method needs less space memory and CPU times than that of ref.\cite{HF-2D}
\bibitem{NM-WI}
For Wigner solutions, $n_m$ is also the number of electrons per unit cell.
\end{thebibliography}
\end{document}